\documentclass[12pt,a4paper,dvips]{article}
\usepackage{amssymb,amsmath,graphicx,a4p}
\usepackage{cite,mcite}
\usepackage{ifthen}
\usepackage{l3_title}
\usepackage{Lep}
\usepackage{epsfig,subfigure,higgs_input}
\journalname{Phys. Lett. B}
\date{November 24, 2000}
\preprint{2000-146}
\Lep{2}

\usepackage[nolists,noheads,nomarkers,tablesfirst]{endfloat}

\begin{document}
\begin{titlepage}
  
  \title{Search for the Standard Model Higgs boson \\
    in \boldmath{\epem} collisions at $\boldsymbol{\rts}$ up to 202\GeV}
  
  \author{L3 Collaboration}
%
%
  \begin{abstract}
    The Standard Model Higgs boson is searched for in 
    233.2\pb of data collected by the L3 detector at  centre of mass
    energies from 192\GeV to  202\GeV.  
    These data are consistent with the expectations
    of Standard Model processes and no evidence of a Higgs signal is
    observed.  A lower limit on the mass of the
    Standard Model Higgs boson of 107.0\GeV is set at the 95\%
    confidence level.
  \end{abstract}

\vspace*{2.cm}
\submitted
\end{titlepage}
%
%
\section{Introduction}
\label{sec:intro}
The Standard Model of the electroweak 
interactions~\cite{sm_glashow} contains a
single Higgs doublet~\cite{higgs_1} which gives rise to
a neutral scalar particle, the Higgs boson. Its mass, \mH, is a free
parameter of the theory. 
A global fit to electroweak precision measurements results in an upper limit 
on \mH of 133\GeV~\cite{l3ewhiggslimit} at 95\% confidence level (CL). 
Results of SM Higgs boson searches in 
\epem collisions were published up to centre of mass
energies of 189\GeV in the Higgs mass range up to 95.3\GeV by 
L3~\cite{l3_smh_189} 
and by other  LEP experiments~\cite{op_smh_189}. 

In this paper, the results of a Higgs search performed on the data
sample collected by L3 at $\rts$ up to 202\GeV are reported.
The dominant Higgs production mode is
\begin{displaymath}
  \epemtoSMHZ \; .
\end{displaymath}
The processes of \WW and \ZZ fusion, which contribute with smaller rate  
to the Higgs
production in the H$\nu\bar\nu$ and H$\rm e^+e^-$ channels, respectively, 
are also considered.
All significant Higgs decay modes are considered in the search. 
The largest sources of background are 
four-fermion final states from \W  and \Z pair production, as well as
\epemtoqqg.

%
\section{Data and Monte Carlo samples}
\label{sec:datamc}
The data were collected using the L3
detector~\cite{l3_1990_1}
at LEP during the year 1999.  
The integrated luminosities are  
29.7\pb at $\rts=$ 191.6\GeV,
83.7\pb at $\rts=$ 195.5\GeV,
82.8\pb at $\rts=$ 199.5\GeV and 
37.0\pb at $\rts=$ 201.8\GeV.

The Higgs production cross sections and branching ratios are calculated
using the HZHA generator~\cite{janotlepii_v2}.
Efficiencies are determined on Monte Carlo samples of Higgs events, generated using
PYTHIA~\cite{PYTHIA}.  
Standard Model background estimates
rely on  the following Monte Carlo programs: 
PYTHIA (\epemtoqqg and $\epem\!\rightarrow\!\Z\epem$),
KORALW~\cite{KORALW} (\epemtoWW), 
KORALZ~\cite{KORALZ}
(\epemtotautau), PHOJET~\cite{PHOJET}
(\epemtoeeqq) and  
EXCALIBUR~\cite{EXCALIBUR} for other four fermion 
final states.
The number of simulated
events for the dominant backgrounds is at least 100
times the number of collected data events for such processes.
For the Higgs signals, at least 2000 events are simulated for  
each search channel and for several masses. Higgs events 
are simulated with \mH between 95 and 110\GeV, with a step of 1\GeV.
Events for Higgs masses between 50 and 95\GeV are simulated with 
a step of 5\GeV.
   
The response of the L3 detector is simulated using the 
GEANT program~\cite{xgeant}, taking into account the effects of multiple
scattering, energy loss and showering in the detector. Hadronic
interactions in the detector are modelled using the GHEISHA
program~\cite{xgheisha}. Time dependent detector inefficiencies,
as monitored during the data taking period, are also simulated.

%
\section{Analysis procedures}
The search for the Standard Model  Higgs boson is based on the study of 
four distinct event topologies
representing approximately 98\% of the \SMHZ decay modes:
$\qqbar\qqbar$, $\qqbar\nnbar$,
$\qqbar\ll\;(\ell=\mathrm{e},\mu,\tau)$ and $\tautau\qqbar$.  With the
exception of the \SMHZtottqq decay mode, 
all the analyses are optimised
for $\bigH\!\rightarrow\!\bbbar$ decay. This mode represents about 82\% of
the Higgs branching fraction in the mass range of interest. 

All these channels are analysed in three stages.
First, 
a high multiplicity hadronic event selection,
preserving most of the Higgs signal,
is applied
to reduce the large background from two-photon processes.  
In a second stage,  
a cut based analysis is applied to the $\qqbar\qqbar$ 
and lepton topologies, while a neural network based 
analysis is used for the $\qqbar\nnbar$ final states.
All the analyses  
use topological and kinematical variables, which are not 
strongly dependent on the Higgs mass.
B hadrons are identified on the basis of an event b-tag 
variable, 
obtained as a combination of the b-tag
for each hadronic jet~\cite{l3_smh_172}. A neural
network is used to calculate the b-tag
for each hadronic jet from the three-dimensional decay lengths,
properties of semileptonic b decays and jet-shape variables.  
The third part of the analysis is the construction of 
a final discriminant for each topology.
This is built from a combination
of the event b-tag variable and the reconstructed Higgs mass, for the cut
based analyses. 
For the  neural network
based analysis, it is a combination of the neural 
network output with the reconstructed Higgs mass.
The distributions of the final discriminants are computed for
the data, for the expected background and signals for each  
Higgs mass hypothesis.

%
\section{The {\boldmath\SMHZtobbqq} selection}

The selection aims to single out events with four jets, two of which
contain b hadrons, while the invariant mass of the other two must be
consistent with the
\Z mass, \mZ. Background from Standard Model processes comes mainly from
\qqbar final states with hard gluons, \WW and
\ZZ events, especially those where one of the \Z bosons decays into b
quarks. 

A preselection to accept high multiplicity hadronic
events is applied. 
The retained events are forced into
four jets with the DURHAM algorithm~\cite{DURHAM}.  
Then a kinematic fit requiring 4-momentum conservation 
is performed. 

In order to differentiate the Higgs signal from
background, selection criteria mainly based on kinematic variables, 
are chosen to maximise
the expected performance of the analysis~\cite{l3_smh_189}.
The quantities used for the selection are: 
the dijet masses,
the minimum jet energy, 
the maximum energy difference between any two jets,
the parameter of the DURHAM scheme for which the 
event is resolved from  three jets
into  four jets, \Ytf, 
and the number of  charged tracks.  
The consistency of the dijet masses with a
given \mH hypothesis is quantified by a 
$\chi^2$-probability that depends on
\mH and \mZ~\cite{l3_smh_189}.  
Only a loose cut is placed on this variable. 

After these cuts, about 85\% of the expected background comes
from \WW events.  These are characterised by their low b-tag values and by
dijet masses close to the W mass.  
The remaining events are then split for the final analysis
into a high purity and a low purity
sample according to the value of the reconstructed dijet mass, \Meq.
This is calculated with
a five constraints  kinematic fit where equal masses are assumed
for the two dijet systems.  
The event is assigned to the high purity
sample if $\Meq > a\times\mH + b $, or to the low purity sample otherwise.
The quantities $a$ and $b$ are
optimised for each centre of mass energy. Typical values 
are $a\sim 0.6$ and $b\sim 40\GeV$.
The low purity sample contains most of the properly reconstructed \WW
events.
Loose b-tag cuts are applied in the final selection and they 
are separately optimised 
at each centre of mass energy for the two samples. 
They are tighter for the low purity one.

The low purity sample contains 
109 events with 108 expected from background, corresponding to
an efficiency of about 28\% for \SMHZtobbqq and \mH=105\GeV. 
The high purity sample contains 32 events with 40 expected from 
background, corresponding to an efficiency of about 30\% for a Higgs mass of
 105\GeV.

\begin{figure}
\begin{center}
\mbox{\epsfxsize=\textwidth \epsffile{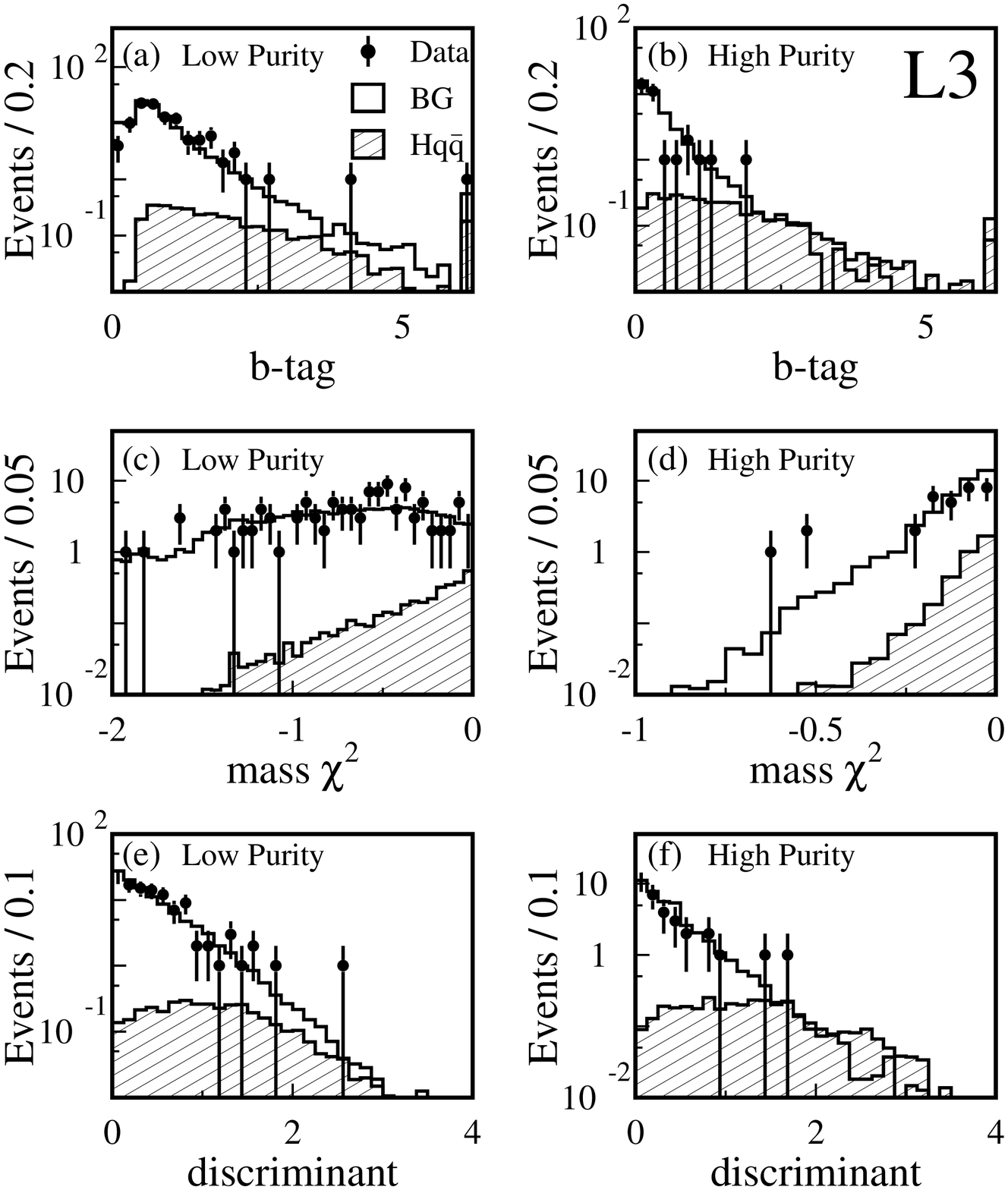}} 
  \caption{The b-tag (a and b), the mass $\chi^2$ (c and d)
   and the final discriminant (e and f) distributions 
   for the low purity and 
   high purity \SMHZtobbqq selections.  The points are the
   192--202\GeV data, the open histograms are the expected  background and
   the hatched histograms are the expected Higgs signal
   from the \SMHZtobbqq channel with $\mH=105\GeV$. 
   The last bin on the right of the b-tag distributions groups 
   the overflows.}
    \label{fig:hqq_btag_final}
  \end{center}
\end{figure}

The final discriminant is then calculated as the weighted
probability~\cite{l3_smh_189} that an event is
consistent with the background distributions of both the b-tag and the
mass variable.  This weighted probability depends
on the mass hypothesis and the discriminant
is calculated for each test mass.
The b-tag, the mass $\chi^2$ and the final discriminant distributions
for all the events in the low purity and high purity samples 
are shown in
Figure~\ref{fig:hqq_btag_final}
for a Higgs mass hypothesis of 105\GeV.
A good agreement between the data and the expected background is 
observed in all the distributions. 

%
\section{The \boldmath\SMHZtobbnn selection}
This selection searches for events with two acoplanar jets containing 
b hadrons, with large missing energy and with  missing mass
consistent with \mZ. 

In the first step, high multiplicity hadronic events are selected.
The events are forced into two jets using the DURHAM algorithm.
The dijet invariant mass must exceed 40\GeV.  
These requirements reduce contributions from
purely leptonic two fermion final states, as well as two-photon
interactions, while retaining a significant fraction of hadronic events
from \epemtoqqg and \W-pair production.  These  backgrounds
are further reduced by requiring the visible mass to be less than
120\GeV and the mass recoiling against the hadronic system to lie
between 50\GeV and 130\GeV.

\begin{figure}[htb]
\begin{center}          
\mbox{\epsfxsize=\textwidth \epsffile{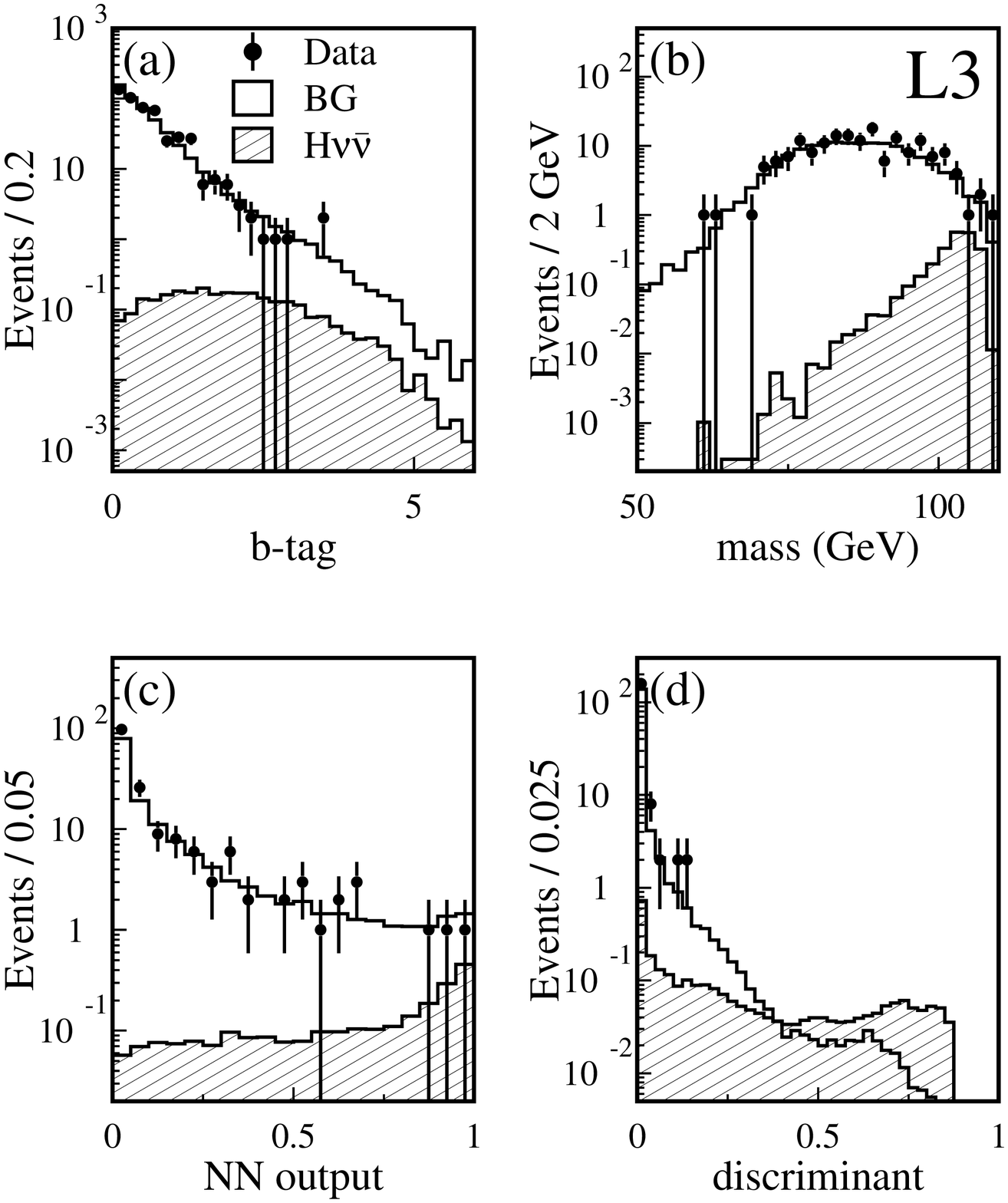}} 
    \caption{Distributions of 
      (a) b-tag, 
      (b) hadronic mass 
      (c) neural network output  and 
      (d) final discriminant for the
      \SMHZtobbnn selection.  The points are the 192--202\GeV data, the open
      histograms are the expected background and the hatched histograms
      the \SMHZtobbnn expected signal with $\mH=105\GeV$.}
\label{fig:hnn_btag_nnet_mass_final} 
\end{center}
\end{figure}

Events from \epemtoqqg are further suppressed by requiring
the longitudinal missing energy to be less than $0.7\rts$, 
the missing energy transverse to the beam axis to
be greater than 5\GeV and the missing momentum vector to be at least
$16^\circ$ away from this axis.  
The energy in the forward luminosity calorimeter is required to be smaller
than 15\GeV.  
The opening angle between the two
jets has to be greater than $69^\circ$ and the angle between the
dijet plane and the beam-axis must be greater than $3^\circ$. 
The b-tag distribution, after the above mentioned cuts
for all centre of mass energies, is shown in
Figure~\ref{fig:hnn_btag_nnet_mass_final}a. 

An additional cut is applied requiring the event b-tag to be larger than 0.5.
After this set of cuts, there are
172 events in the data, while 149 are expected from background 
processes. The efficiency for \SMHZtobbnn with
$\mH=105\GeV$ is 62\%.

A kinematic fit imposing
4-momentum conservation and requiring the missing mass to be \mZ is
performed to compute the hadronic mass. The distribution of this variable
is shown in  Figure~\ref{fig:hnn_btag_nnet_mass_final}b.
The output of a mass independent neural network~\cite{l3_smh_189} 
is then combined with the hadronic mass to
build the final discriminant.
The distribution of the neural network output 
is shown in Figure~\ref{fig:hnn_btag_nnet_mass_final}c. 
The final discriminant is presented in
Figure~\ref{fig:hnn_btag_nnet_mass_final}d for the mass hypothesis
$\mH=105\GeV$.  The observed data in the \SMHZtobbnn analysis are
compatible with the background expectations.

%
\section{The \boldmath\SMHZtobbee and \boldmath\SMHZtobbmm selections}
The signatures for the \SMHZtobbee and \SMHZtobbmm processes 
are a pair of high energy electrons or muons, with an invariant 
mass compatible with \mZ and two hadronic jets with b quark content.

A high multiplicity selection is applied, also  requiring 
two well identified electrons or
muons.  The visible energy must be larger than $0.7\rts$ for the
electron analysis and $0.4\rts$ for the muon analysis.  In the
\SMHZtobbee analysis, the lepton pair must have an opening angle
greater than $100^\circ$, reduced to $90^\circ$ in 
the \SMHZtobbmm case. 
Moreover, in the \SMHZtobbee 
channel the opening angle between the two jets must be at least $50^\circ$. 
The ratio between the transverse missing momentum and the 
visible energy should be less than 0.2 in the \SMHZtobbee channel and less than
0.4 in the \SMHZtobbmm channel.
The value of \Ytf must be larger than 0.0009.
Finally, the invariant mass of the leptons after a kinematic
fit imposing 4-momentum conservation must be between 60\GeV and
110\GeV for the electrons and 50\GeV and 125\GeV for the muons.  

In the electron channel, 22 events are selected 
with 20.2 expected from the background and with a signal efficiency 
of 76\% for a Higgs signal of 105\GeV.  In the muon channel there are 13
events with 9.2 expected from the background,
with a signal efficiency of 56\%.

\begin{figure}[htb]
  \begin{center}
\mbox{\epsfxsize=\textwidth\epsffile{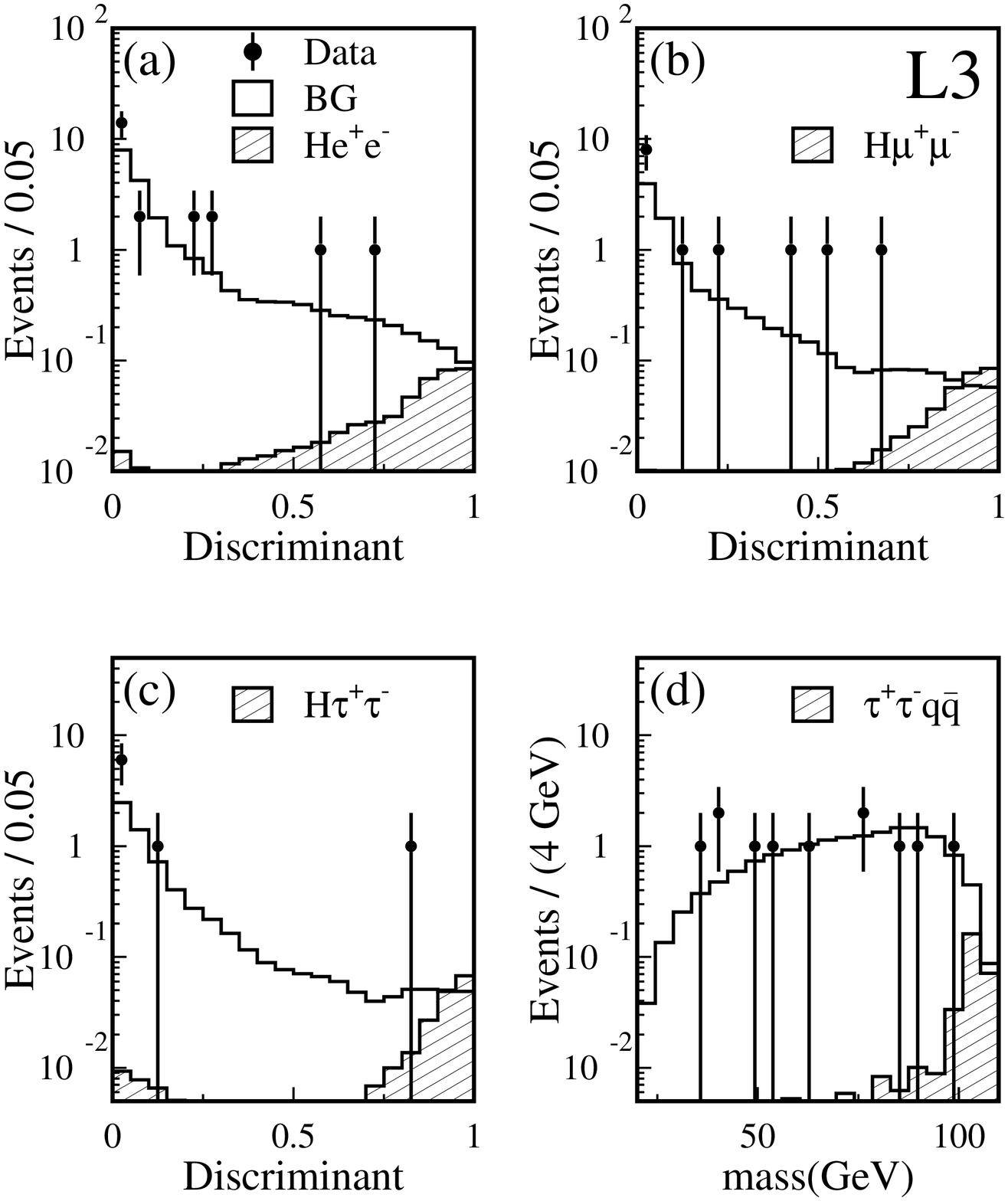}} 
\caption{Distributions of the final discriminant for the (a) \SMHZtobbee,
      (b) \SMHZtobbmm, (c) \SMHZtobbtt and (d) \SMHZtottqq selections.
      The 192--202\GeV data (points) are compared to the 
expected background (open histogram)  and to the expected Higgs signal 
(hatched histogram) of 105\GeV.
      The signal events
      in the \SMHZtobbtt and \SMHZtottqq distributions include the
      cross-efficiencies for these channels.
      Events are uniquely assigned to only one of these channels.}
    \label{fig:hll_final}
  \end{center}
\end{figure}

A kinematic fit that requires 4-momentum conservation and
constrains the mass of the leptons to \mZ is performed.
The dijet mass after this fit is combined with the 
b-tag values  of the two jets, to form 
the final discriminant~\cite{l3_smh_189}. 
The distributions of the discriminant for the electron
and muon channels are shown in   
Figure~\ref{fig:hll_final}a and ~\ref{fig:hll_final}b respectively, 
for the data compared to the expected background and 
a 105\GeV Higgs signal.  The data are consistent with the 
background predictions.

%
\section{The \boldmath\SMHZtobbtt and \boldmath\SMHZtottqq selections}
The \SMHZtobbtt and \SMHZtottqq processes result in 
similar final states, partially 
distinguished by mass and b-tag information.  The semileptonic
W and Z pair decays constitute the most significant background sources.

Two inclusive selections are performed, one based on tau
identification, the particle-based selection, and the other 
relying more on the
event kinematics, the jet-based selection. Events are accepted if they
pass either of the two selections.

First, a high multiplicity selection is applied, 
also requiring a visible energy greater than
$0.4\rts$.  The events are required to have a value of \Ytf 
larger than 0.0025.  Background from
\epemtoqqg is reduced by rejecting events containing
photons with energies greater than 40\GeV.  The contribution of
$\WW\!\rightarrow\!\qqbar\ell\nu \; (\ell=\e,\mu)$ is reduced by
requiring the energy of electrons and muons to be below 40\GeV.

In the particle-based selection, tau leptons are identified either by their
decay into electrons or muons, or as an isolated low-multiplicity jet
with 1 or 3 tracks and unit charge.  In the jet-based selection, the
event is forced into four jets using the DURHAM algorithm. Two of the
jets must have less than 4 tracks each. These jets are considered as
tau candidates, but at least one of them must coincide within a $3^\circ$
cone with a tau
candidate identified by the particle-based selection. 
Both tau candidates must be separated from the hadronic jets by at least
25$^\circ$. Background contamination from fully hadronic W pair decays is
reduced by rejecting events where both tau candidates decay into 3 charged
particles and by requiring the visible energy to be smaller than
$0.95\rts$ for the particle-based selection and smaller than $0.9\rts$ for the
jet-based one.  Moreover, in the jet-based selection, the 
missing momentum vector should be at least 18$^\circ$ from the beam axis, 
in order to reduce the $\qqbar(\gamma)$ contamination.

The invariant masses of the ditau and the dijet are
obtained from a kinematic fit which imposes 4-momentum conservation.
An event qualifies for the \SMHZtobbtt channel if the ditau  mass
is consistent with the mass of the Z boson by
lying between 70\GeV and 125\GeV.  Similarly, an event is assigned to
the \SMHZtottqq channel if the dijet mass fulfils the same
requirement.  The opening angle of the taus or jets
assigned to the Higgs boson must be larger than $70^\circ$. Those
assigned to the \Z must be at least $100^\circ$ apart.
Events selected by the \SMHZtobbee and the \SMHZtobbmm analyses amount 
to 3\% and are not considered here.

In total, 19 events pass
either of the tau selections, with 22.3 events expected 
from background processes. The efficiency is 
29\% for both \SMHZtobbtt and \SMHZtottqq at $\mH=105\GeV$.

The discriminant is defined
as in the \SMHZtobbee and \SMHZtobbmm analyses.
Events passing both \SMHZtobbtt and \SMHZtottqq selections are 
assigned to the channel for which the 
value of this discriminant is higher. This discriminant is used as 
the final variable for the  \SMHZtobbtt selection. 
For the \SMHZtottqq selection, the mass
of the tau pair, calculated by constraining the invariant mass of
the two other jets to \mZ, is used as the final discriminant.  
Distributions of the final
discriminants are shown in the Figures~\ref{fig:hll_final}c 
and~\ref{fig:hll_final}d
for data, 
expected background and a 105\GeV Higgs
signal.  No evidence of a signal appears in either of the tau
channels. 

%
\section{Combined results}
\label{sec:results}
\begin{table}[htbp]
\small
  \begin{center}
    \begin{tabular}{lllrrrrrrrr}
      \hline
      $\rts=192\GeV$ &&&
\multicolumn{8}{c}{Mass hypothesis\rule{0em}{1.2em}} \\
      \multicolumn{2}{c}{Selection} && \multicolumn{3}{c}{$\mH=100\GeV$} && \multicolumn{3}{c}{$\mH=105\GeV$} &  \\
      \cline{1-2} \cline{4-11}
      \bigH & \Z && $\rm N_D$ & $\rm N_B$ & \multicolumn{2}{l}{$\rm N_S$} &$\rm N_D$  & $\rm N_B$ & \multicolumn{2}{l}{$\rm N_S$}   \\
      \hline
\bbbar\rule{0em}{1.2em}
       & \qqbar && 4   & 2.9  & 0.5  &&  0 & 0.2 & 0.0  &\\
\bbbar & \nnbar && 1   & 1.8  & 0.2  &&  0 & 0.1 & 0.0  &\\
\bbbar & \epem  && 0   & 0.1  & 0.1  &&  0 & 0.0 & 0.0  &\\
\bbbar & \mumu  && 1   & 0.1  & 0.0  &&  0 & 0.1 & 0.0  &\\
\bbbar & \tautau&& 0   & 0.0  & 0.0  &&  0 & 0.0 & 0.0  &\\
\tautau& \qqbar && 0   & 0.1  & 0.0  &&  0 & 0.0 & 0.0  &\\
\hline \multicolumn{2}{l}{Total}
                && 6  & 5.0  & 0.8   &&  0 & 0.4 & 0.0   & \\
      \hline\hline
      $\rts=196\GeV$&&& \multicolumn{8}{c}{Mass hypothesis\rule{0em}{1.2em}} \\
      \multicolumn{2}{c}{Selection} && \multicolumn{3}{c}{$\mH=100\GeV$} && \multicolumn{3}{c}{$\mH=105\GeV$} &  \\
      \cline{1-2} \cline{4-11}
      \bigH & \Z && $\rm N_D$ & $\rm N_B$ & \multicolumn{2}{l}{$\rm N_S$} &$\rm N_D$  & $\rm N_B$ & \multicolumn{2}{l}{$\rm N_S$}   \\
      \hline
\bbbar\rule{0em}{1.2em}
       & \qqbar && 31 & 28.2  & 5.2  &&  6 &  5.2 & 0.7 &\\
\bbbar & \nnbar &&  4 &  6.9  & 1.7  &&  0 &  1.3 & 0.2 &\\
\bbbar & \epem  &&  1 &  0.7  & 0.4  &&  0 &  0.2 & 0.0 &\\
\bbbar & \mumu  &&  1 &  0.6  & 0.3  &&  1 &  0.2 & 0.0 &\\
\bbbar & \tautau&&  0 &  0.2  & 0.1  &&  1 &  0.1 & 0.0 &\\
\tautau& \qqbar &&  1 &  0.3  & 0.3  &&  0 &  0.1 & 0.0 &\\ \hline \multicolumn{2}{l}{Total}
                && 38 & 36.9  & 8.0 &&  9 &  7.1 & 0.9 &\\
      \hline
      \hline
      $\rts=200\GeV$&&& \multicolumn{8}{c}{Mass hypothesis\rule{0em}{1.2em}} \\
      \multicolumn{2}{c}{Selection} && \multicolumn{3}{c}{$\mH=100\GeV$} && \multicolumn{3}{c}{$\mH=105\GeV$} &  \\
      \cline{1-2} \cline{4-11}
      \bigH & \Z && $\rm N_D$ & $\rm N_B$ & \multicolumn{2}{l}{$\rm N_S$} &$\rm N_D$  & $\rm N_B$ & \multicolumn{2}{l}{$\rm N_S$}   \\
      \hline
\bbbar\rule{0em}{1.2em}
       & \qqbar &&   24 & 29.9 & 6.9 &&  18  & 18.2 &3.6 &\\
\bbbar & \nnbar &&   13 &  8.6 & 2.2 &&   4  &  3.6 &1.1 &\\
\bbbar & \epem  &&    3 &  1.2 & 0.5 &&   1  &  1.0 &0.2 &\\
\bbbar & \mumu  &&    0 &  0.8 & 0.4 &&   0  &  0.4 &0.2 &\\
\bbbar & \tautau&&    0 &  0.5 & 0.2 &&   0  &  0.2 &0.1 &\\
\tautau& \qqbar &&    0 &  0.9 & 0.4 &&   0  &  0.5 &0.1 &\\ \hline \multicolumn{2}{l}{Total}
                && 40   & 41.9 & 10.6 && 23  & 23.9 & 5.3 &\\
      \hline
      \hline
      $\rts=202\GeV$&&& \multicolumn{8}{c}{Mass hypothesis\rule{0em}{1.2em}} \\
      \multicolumn{2}{c}{Selection} && \multicolumn{3}{c}{$\mH=100\GeV$} && \multicolumn{3}{c}{$\mH=105\GeV$} &  \\
      \cline{1-2} \cline{4-11}
      \bigH & \Z && $\rm N_D$ & $\rm N_B$ & \multicolumn{2}{l}{$\rm N_S$} &$\rm N_D$  & $\rm N_B$ & \multicolumn{2}{l}{$\rm N_S$}   \\
      \hline
\bbbar\rule{0em}{1.2em}
       & \qqbar && 13 & 12.5  & 3.0 &&  7 &  8.5 &  2.0  &\\
\bbbar & \nnbar &&  6 &  2.5  & 1.0 &&  4 &  2.7 &  0.7  &\\
\bbbar & \epem  &&  0 &  0.5  & 0.2 &&  0 &  0.6 &  0.1  &\\
\bbbar & \mumu  &&  0 &  0.4  & 0.2 &&  0 &  0.2 &  0.1  &\\
\bbbar & \tautau&&  0 &  0.2  & 0.1 &&  0 &  0.1 &  0.1  &\\
\tautau& \qqbar &&  0 &  0.4  & 0.2 &&  0 &  0.3 &  0.1  &\\ \hline \multicolumn{2}{l}{Total}
                && 19 & 16.5  & 4.7 && 11 & 12.4 &  3.1 &\\
      \hline
    \end{tabular}
    \caption{The number of observed candidates ($\rm N_D$), 
 expected background events ($\rm N_B$) and 
expected signal ($\rm N_S$) for the $\rts=192\GeV,196\GeV,200\GeV,202\GeV$
      data after a
      cut on the final discriminant corresponding to a
      signal-to-background ratio greater than 0.05.  
This cut is used to calculate the confidence levels.}\vspace{0.5em}
    \label{tab:results}
  \end{center}
\end{table}
\normalsize
The results of all the previously described analyses are combined together
to set a lower limit
on the mass of the Standard Model Higgs boson.
A combined CL on the absence of a signal is derived from
the distributions of final discriminants in a scan
over \mH from 50\GeV to 110\GeV.  The CL is calculated using the
technique of References~\cite{l3_smh_172,new_method}, 
which takes into account the correlated and the statistical errors.

The systematic uncertainties on the signal and background expectations are
derived using the same procedure adopted in 
previous Standard Model Higgs
searches~\cite{l3_smh_172}.  The overall systematic
uncertainty is estimated to be  10\% on the number of background events 
and 4\% on the number of signal events.  The statistical uncertainty on the
background arising from the finite number of generated Monte Carlo events is
uncorrelated from bin to bin in the final discriminant
distributions, and has little effect on the CL.  Bins with a
signal-to-background ratio below 0.05 are not considered
in the calculation of the CL.  This cut was chosen to maximise the
median CL, as calculated from a large number of Monte Carlo
experiments, thereby minimising the degradation of the result due to
these systematic and statistical uncertainties.
The results of all the analyses after such
a signal-to-background cut are summarised in Table~\ref{tab:results}
for the data, the expected background and for  
Higgs signals of 100\GeV and 105\GeV.  The number of signal
events includes cross-efficiencies from other channels, fusion
processes and charm and gluonic Higgs decays.

\begin{figure}[htbp]
  \begin{center}
\mbox{\epsfxsize=0.95\textwidth \epsffile{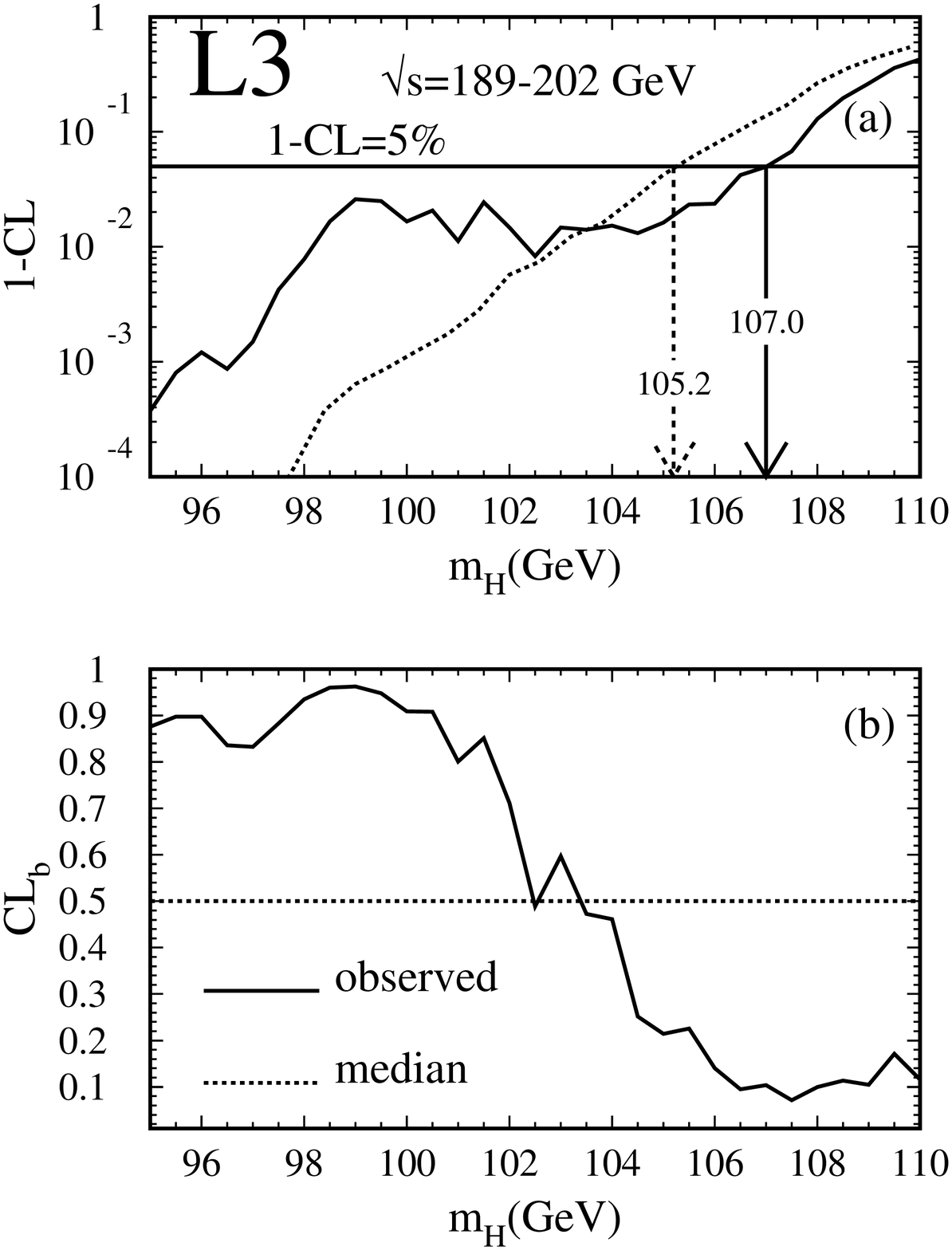}} 
\vspace*{-1.cm}     \caption{(a) The observed (solid line)
 and expected median  (dashed line) 
signal confidence levels (1-CL) as
      a function of the Higgs mass, (b) the observed (solid line)
 and expected (dashed line) background
      confidence level CL$_b$ as a function of the Higgs mass.
      The lower limit on the Higgs mass is
      set at 107.0\GeV at the 95\% CL.}
    \label{fig:limit}
  \end{center}
\end{figure}

The measured value of the CL as a function of the Standard Model Higgs
boson mass in the range $95\leq\mH\leq 110\GeV$ is shown in
the Figure~\ref{fig:limit}a, along with the median of the CL
distribution as calculated from a large sample of Monte Carlo
experiments under the background-only hypothesis. 
The median CL represents
the sensitivity of the analysis and is equal to 95\% at
$\mH=105.2\GeV$.
The results of L3 Standard Model
Higgs searches at lower centre of mass
energies~\cite{l3_smh_189} are included in
the calculation of these confidence levels.  Values of \mH from 50\GeV
to 95\GeV are excluded in the Standard Model 
with  a confidence level greater than 99\%.

Figure~\ref{fig:limit}b shows the background confidence level, CL$_b$.
The observed CL$_b$ is the probability to observe a smaller 
number of events than
the one actually observed, for a background-only hypothesis. 
The expected CL$_b$  for a background-only hypothesis is 0.5. 
The  observed CL$_b$ gives a  measure of the consistency of 
the data with the expected background. 
The relatively high value of CL$_b$ in the mass region between 
96 and 100\GeV reflects 
a 2.2 sigma excess of data relative to the background 
predictions in this mass region. This excess is mainly due to few
candidates in the $\boldmath\SMHZtobbqq$ channel and to one candidate
in the $\boldmath\SMHZtobbqq$ channel with a reconstructed Higgs mass
of 100.5\GeV. For \mH=107\GeV,
where the observed CL falls below 95\%, the  
CL$_b$ is 8\%.

The lower limit on the Standard Model Higgs boson mass is set at
\begin{displaymath}
  \mH > 107.0\GeV \; \mathrm{at\ 95\% \ CL.}
\end{displaymath}
This new lower limit improves upon and supersedes our previously
published results. 
A similar result from the 192--202\GeV data was reported~\cite{al_smh_200}.
Results from the year 2000 LEP run were recently 
published~\cite{al_smh_207,l3_smh_207}. 
%
%
\section*{Acknowledgements}
We acknowledge the efforts of the engineers and technicians who have
participated in the construction and maintenance of L3 and express our
gratitude to the CERN accelerator divisions for the superb performance
of LEP.
\newpage

\bibliographystyle{l3stylem}

\newpage
\typeout{   }     
\typeout{Using author list for paper 229 -- ? }
\typeout{$Modified: Nov 18 2000 by smele $}
\typeout{!!!!  This should only be used with document option a4p!!!!}
\typeout{   }
%
%
%
%
%
%

\newcount\tutecount  \tutecount=0
\def\tutenum#1{\global\advance\tutecount by 1 \xdef#1{\the\tutecount}}
\def\tute#1{$^{#1}$}
\tutenum\aachen            
\tutenum\nikhef            
\tutenum\mich              
\tutenum\lapp              
\tutenum\basel             
\tutenum\lsu               
\tutenum\beijing           
\tutenum\berlin            
\tutenum\bologna           
\tutenum\tata              
\tutenum\ne                
\tutenum\bucharest         
\tutenum\budapest          
\tutenum\mit               
\tutenum\debrecen          
\tutenum\florence          
\tutenum\cern              
\tutenum\wl                
\tutenum\geneva            
\tutenum\hefei             
\tutenum\seft              
\tutenum\lausanne          
\tutenum\lecce             
\tutenum\lyon              
\tutenum\madrid            
\tutenum\milan             
\tutenum\moscow            
\tutenum\naples            
\tutenum\cyprus            
\tutenum\nymegen           
\tutenum\caltech           
\tutenum\perugia           
\tutenum\cmu               
\tutenum\prince            
\tutenum\rome              
\tutenum\peters            
\tutenum\potenza           
\tutenum\riverside         
\tutenum\salerno           
\tutenum\ucsd              
\tutenum\sofia             
\tutenum\korea             
\tutenum\alabama           
\tutenum\utrecht           
\tutenum\purdue            
\tutenum\psinst            
\tutenum\zeuthen           
\tutenum\eth               
\tutenum\hamburg           
\tutenum\taiwan            
\tutenum\tsinghua          

{
\parskip=0pt
\noindent
{\bf The L3 Collaboration:}
\ifx\selectfont\undefined
 \baselineskip=10.8pt
 \baselineskip\baselinestretch\baselineskip
 \normalbaselineskip\baselineskip
 \ixpt
\else
 \fontsize{9}{10.8pt}\selectfont
\fi
\medskip
\tolerance=10000
\hbadness=5000
\raggedright
\hsize=162truemm\hoffset=0mm
\def\r{\rlap,}
\noindent

M.Acciarri\r\tute\milan\
P.Achard\r\tute\geneva\ 
O.Adriani\r\tute{\florence}\ 
M.Aguilar-Benitez\r\tute\madrid\ 
J.Alcaraz\r\tute\madrid\ 
G.Alemanni\r\tute\lausanne\
J.Allaby\r\tute\cern\
A.Aloisio\r\tute\naples\ 
M.G.Alviggi\r\tute\naples\
G.Ambrosi\r\tute\geneva\
% ...
\newpage
\begin{list}{A}{\itemsep=0pt plus 0pt minus 0pt\parsep=0pt plus 0pt minus 0pt
                \topsep=0pt plus 0pt minus 0pt}
\item[\aachen]
 I. Physikalisches Institut, RWTH, D-52056 Aachen, FRG$^{\S}$\\
 III. Physikalisches Institut, RWTH, D-52056 Aachen, FRG$^{\S}$
\item[\nikhef] National Institute for High Energy Physics, NIKHEF, 
     and University of Amsterdam, NL-1009 DB Amsterdam, The Netherlands
\item[\mich] University of Michigan, Ann Arbor, MI 48109, USA
\item[\lapp] Laboratoire d'Annecy-le-Vieux de Physique des Particules, 
     LAPP,IN2P3-CNRS, BP 110, F-74941 Annecy-le-Vieux CEDEX, France
\item[\basel] Institute of Physics, University of Basel, CH-4056 Basel,
     Switzerland
\item[\lsu] Louisiana State University, Baton Rouge, LA 70803, USA
\item[\beijing] Institute of High Energy Physics, IHEP, 
  100039 Beijing, China$^{\triangle}$ 
\item[\berlin] Humboldt University, D-10099 Berlin, FRG$^{\S}$
\item[\bologna] University of Bologna and INFN-Sezione di Bologna, 
     I-40126 Bologna, Italy
\item[\tata] Tata Institute of Fundamental Research, Bombay 400 005, India
\item[\ne] Northeastern University, Boston, MA 02115, USA
\item[\bucharest] Institute of Atomic Physics and University of Bucharest,
     R-76900 Bucharest, Romania
\item[\budapest] Central Research Institute for Physics of the 
     Hungarian Academy of Sciences, H-1525 Budapest 114, Hungary$^{\ddag}$
\item[\mit] Massachusetts Institute of Technology, Cambridge, MA 02139, USA
\item[\debrecen] KLTE-ATOMKI, H-4010 Debrecen, Hungary$^\P$
\item[\florence] INFN Sezione di Firenze and University of Florence, 
     I-50125 Florence, Italy
\item[\cern] European Laboratory for Particle Physics, CERN, 
     CH-1211 Geneva 23, Switzerland
\item[\wl] World Laboratory, FBLJA  Project, CH-1211 Geneva 23, Switzerland
\item[\geneva] University of Geneva, CH-1211 Geneva 4, Switzerland
\item[\hefei] Chinese University of Science and Technology, USTC,
      Hefei, Anhui 230 029, China$^{\triangle}$
\item[\lausanne] University of Lausanne, CH-1015 Lausanne, Switzerland
\item[\lecce] INFN-Sezione di Lecce and Universit\`a Degli Studi di Lecce,
     I-73100 Lecce, Italy
\item[\lyon] Institut de Physique Nucl\'eaire de Lyon, 
     IN2P3-CNRS,Universit\'e Claude Bernard, 
     F-69622 Villeurbanne, France
\item[\madrid] Centro de Investigaciones Energ{\'e}ticas, 
     Medioambientales y Tecnolog{\'\i}cas, CIEMAT, E-28040 Madrid,
     Spain${\flat}$ 
\item[\milan] INFN-Sezione di Milano, I-20133 Milan, Italy
\item[\moscow] Institute of Theoretical and Experimental Physics, ITEP, 
     Moscow, Russia
\item[\naples] INFN-Sezione di Napoli and University of Naples, 
     I-80125 Naples, Italy
\item[\cyprus] Department of Natural Sciences, University of Cyprus,
     Nicosia, Cyprus
\item[\nymegen] University of Nijmegen and NIKHEF, 
     NL-6525 ED Nijmegen, The Netherlands
\item[\caltech] California Institute of Technology, Pasadena, CA 91125, USA
\item[\perugia] INFN-Sezione di Perugia and Universit\`a Degli 
     Studi di Perugia, I-06100 Perugia, Italy   
\item[\cmu] Carnegie Mellon University, Pittsburgh, PA 15213, USA
\item[\prince] Princeton University, Princeton, NJ 08544, USA
\item[\rome] INFN-Sezione di Roma and University of Rome, ``La Sapienza",
     I-00185 Rome, Italy
\item[\peters] Nuclear Physics Institute, St. Petersburg, Russia
\item[\potenza] INFN-Sezione di Napoli and University of Potenza, 
     I-85100 Potenza, Italy
\item[\riverside] University of Californa, Riverside, CA 92521, USA
\item[\salerno] University and INFN, Salerno, I-84100 Salerno, Italy
\item[\ucsd] University of California, San Diego, CA 92093, USA
\item[\sofia] Bulgarian Academy of Sciences, Central Lab.~of 
     Mechatronics and Instrumentation, BU-1113 Sofia, Bulgaria
\item[\korea]  Laboratory of High Energy Physics, 
     Kyungpook National University, 702-701 Taegu, Republic of Korea
\item[\alabama] University of Alabama, Tuscaloosa, AL 35486, USA
\item[\utrecht] Utrecht University and NIKHEF, NL-3584 CB Utrecht, 
     The Netherlands
\item[\purdue] Purdue University, West Lafayette, IN 47907, USA
\item[\psinst] Paul Scherrer Institut, PSI, CH-5232 Villigen, Switzerland
\item[\zeuthen] DESY, D-15738 Zeuthen, 
     FRG
\item[\eth] Eidgen\"ossische Technische Hochschule, ETH Z\"urich,
     CH-8093 Z\"urich, Switzerland
\item[\hamburg] University of Hamburg, D-22761 Hamburg, FRG
\item[\taiwan] National Central University, Chung-Li, Taiwan, China
\item[\tsinghua] Department of Physics, National Tsing Hua University,
      Taiwan, China
\item[\S]  Supported by the German Bundesministerium 
        f\"ur Bildung, Wissenschaft, Forschung und Technologie
\item[\ddag] Supported by the Hungarian OTKA fund under contract
numbers T019181, F023259 and T024011.
\item[\P] Also supported by the Hungarian OTKA fund under contract
  numbers T22238 and T026178.
\item[$\flat$] Supported also by the Comisi\'on Interministerial de Ciencia y 
        Tecnolog{\'\i}a.
\item[$\sharp$] Also supported by CONICET and Universidad Nacional de La Plata,
        CC 67, 1900 La Plata, Argentina.
\item[$\diamondsuit$] Also supported by Panjab University, Chandigarh-160014, 
        India.
\item[$\triangle$] Supported by the National Natural Science
  Foundation of China.
\end{list}
}
\vfill


\newpage

\end{document}